\documentclass[twocolumn]{aastex63}

\graphicspath{{./}{figs/}}

\usepackage{amsmath}
\usepackage{appendix}
\usepackage{needspace}

\usepackage{tikz}
\usetikzlibrary{decorations.pathmorphing, patterns, decorations.text}
\definecolor{customRed}{HTML}{CA0020}
\definecolor{customBlue}{HTML}{0571b0}

\received{April 1, 2024}%\today}
\shorttitle{SLAB Inspirals and Hubble Tensions}
\shortauthors{Bellinger, Stegmann \& Wagg}

\begin{document}
\title{Stupendously Large Black Hole Coalescence and Hubble Tensions}

\correspondingauthor{E.\ P.\ Bellinger}
\email{earl.bellinger@yale.edu}

\author[0000-0003-4456-4863]{Earl Patrick Bellinger} 
\affil{Department of Astronomy, Yale University, CT, USA} 

\author[0000-0003-2340-8140]{Jakob Stegmann}
\affiliation{Max-Planck-Institut für Astrophysik, 
%Karl-Schwarzschild-Str. 1, 85741 
Garching bei München, Germany}

\author[0000-0001-6147-5761]{Tom Wagg} 
\affil{Department of Astronomy, University of Washington, Seattle, WA, USA}

\begin{abstract} 
Local and distant measurements of the Hubble constant are in significant tension: local measurements of the Hubble constant appear to show a Universe that is significantly contracted when compared to distant measurements. 
From the point of view of an observer, a passing gravitational wave could cause the Universe to appear locally contracted and expanded in a quadrupolar pattern. 
The inspiral of a pair of stupendously large black holes (SLABs) with a chirp mass of $\sim10^{21}~\rm{M}_\odot$ may produce gravitational radiation with sufficiently large amplitude and wavelength to increase $H_0$ in one direction, and decrease it by the same amount in the other. 
These tensions would then oscillate with a period corresponding to half the orbital period of the binary. 
If such a gravitational wave were aligned with the plane of the Milky Way, most readily visible galaxies would appear closer than they actually are, thereby causing the apparent Hubble tension. 
Due to the long binary period, we would be in the same phase of the gravitational wave for the complete history of astronomical observation. 
The opposite tension would be visible in the orthogonal directions, thus giving the opportunity to falsify the existence of inspiraling SLAB binaries. 
As a corollary, the Hubble tension may place an upper limit on the maximum mass of inspiraling black holes in the Universe. 
\end{abstract}


\section{Introduction} \label{sec:intro} 
The tension between local and cosmological measurements of the Hubble constant ($H_0$) stands as one of the most intriguing puzzles in contemporary astrophysics. Local observations based on Cepheid-host galaxies suggest a value of $H_0$ that is significantly higher than that derived from the Cosmic Microwave Background (CMB) and other cosmological probes. This discrepancy, known as the Hubble tension, implies a fundamental gap in our understanding of the Universe's expansion rate and possibly its underlying physical laws.

Recent theoretical work has explored a wide array of potential resolutions to this tension, ranging from new physical phenomena to systematic errors in observational methods \citep{2022JHEAp..34...49A}. However, many proposed solutions operate within the framework of known physics and observable phenomena. Here we entertain a less conventional hypothesis: the role of stupendously large black hole \citep[SLABs,][]{2021MNRAS.501.2029C} binaries in shaping our observation of the Universe.

The concept of SLABs, black holes with masses above $10^{10}~\mathrm{M}_\odot$, extends beyond current observational evidence and theoretical models. These entities, if they exist, would represent extreme endpoints of black hole growth. They would grow primarily through mergers rather than accretion, due to the unfeasibility of sustaining an accretion disk above a certain mass threshold \citep{2009MNRAS.393..838N, 2016MNRAS.456L.109K}. The upper mass limit of such black holes might only be constrained by the dynamics of the observable universe itself, with some estimates of the upper mass limit of any such black hole anywhere in the Universe reaching as high as $10^{20}~\mathrm{M}_\odot$ \citep{2021MNRAS.501.2029C}. It is worth noting that this figure is far more massive than the most massive galaxies ($\sim 10^{12}~\mathrm{M}_\odot$) and galaxy clusters ($\sim 10^{15}~\mathrm{M}_\odot$). 

Here we explore an imaginative idea: that the inspiral of two supermassive SLABs could produce gravitational waves (GWs) of such magnitude and wavelength that they significantly alter our observation of $H_0$. According to General Relativity, GWs distort spacetime in a quadrupolar pattern, locally contracting and expanding space as they pass. An extraordinarily large and long-period GW, emanating from a distant SLAB binary, could thus impact the measured distances to galaxies, depending on the observer's alignment with the wave's propagation direction.

%In particular, if such a wave were oriented along the plane of the Milky Way, it would selectively influence measurements of H0, making distant galaxies appear closer and hence increasing the locally measured value of the Hubble constant.
In particular, if such a wave were oriented along the plane of the Milky Way, it would selectively influence measurements of $H_0$, making some distant galaxies in directions orthogonal to the wave's propagation to appear closer and hence increasing the inferred value of the Hubble constant. 
Due to the long orbital period of SLAB binaries, we remain in one phase of the wave over the entire history of astronomical observation. 
Cepheid-host galaxies are difficult to observe near the galactic plane due to extinction, and thus are so far only observed at relatively high galactic latitudes \citep{2022ApJ...934L...7R}, and hence currently observed galaxies could be probing the contracting portion of the wave. 
This theory offers a falsifiable prediction: galaxies observed in the expanding portion of the gravitational wave should exhibit an opposite tension, a hypothesis that could be tested with future observations. 

This scenario also invites consideration of the constraints that the Hubble tension places on the maximum mass of merging black holes within the observable universe. If the merger of SLABs can influence $H_0$ measurements to the observed degree, then the existence of such massive black holes may be limited by the need to preserve the consistency of cosmological measurements. 

\section{Observations}
\begin{figure*}
    \centering
    \includegraphics[width=0.9\textwidth]{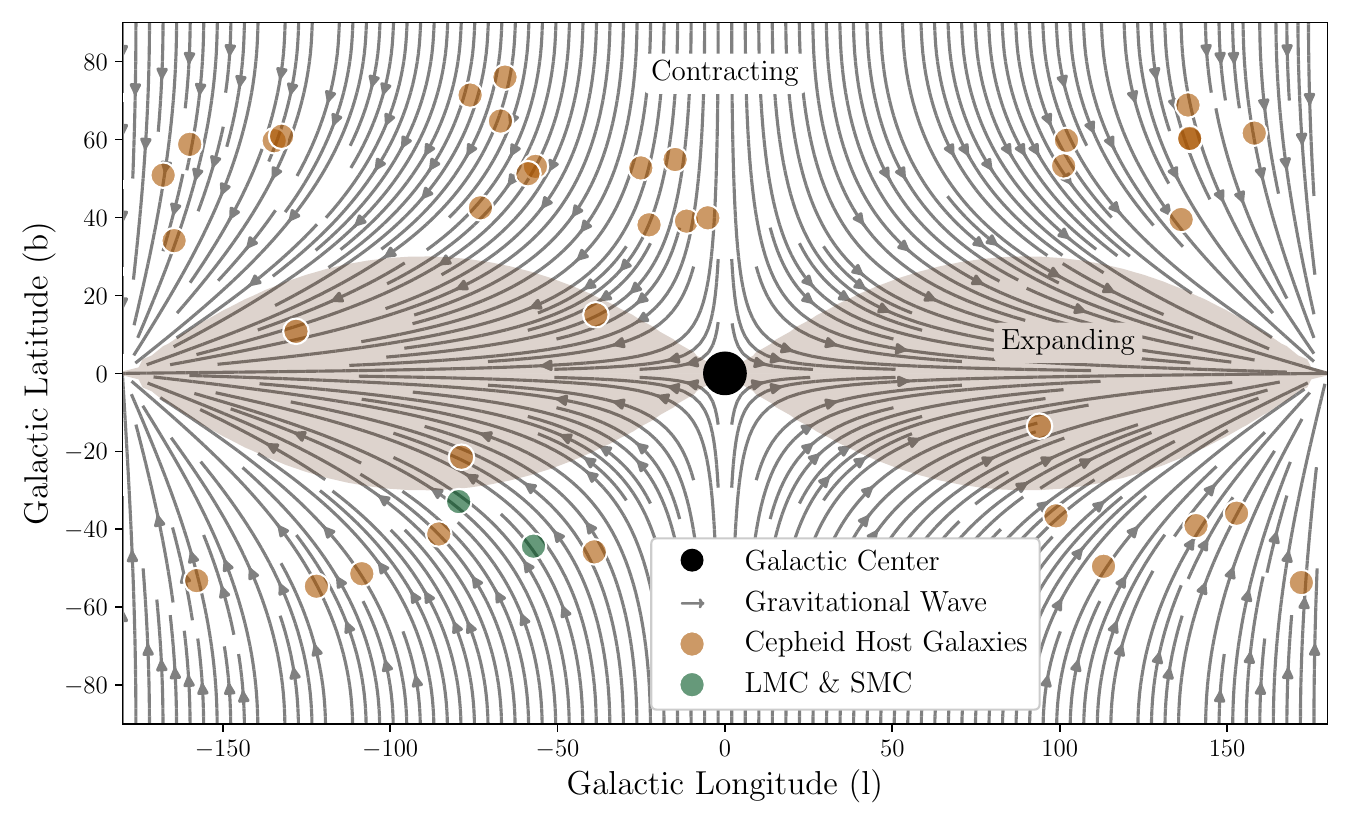}
    \caption{Artistic depiction of a large-amplitude gravitational wave (arrows) aligned with the galactic center (black dot) affecting the distances of Cepheid host galaxies used to calibrate the Hubble constant (orange and green points). Only four galaxies are in the primarily expanding portion of the gravitational wave (shaded region); the rest are in the contracting portion. \label{fig:slab-cepheids}}
\end{figure*}

Here we aim to illustrate that the current observations of Cepheid host galaxies could in principle be biased by a large-amplitude gravitational wave. 
We obtained the positions of the 40 galaxies used to calibrate the local Hubble constant from \citet{2022ApJ...934L...7R}; these are shown in Figure~\ref{fig:slab-cepheids}. 

For a simplistic artistic illustration of a quadrupolar wave, we adopt the following force field: 
\begin{align}
U = &\sin\left(\text{rad}(l)\right) \cdot \cos^2\left(\text{rad}(b)\right) \\
V = -&\sin\left(\text{rad}(b)\right) \cdot \cos\left(\text{rad}(b)\right)
\end{align}
where $l$ is the galactic longitude, $b$ is the galactic latitude, $U$ is the velocity component in the direction of the Galactic center, and 
$V$ is the velocity component in the direction of the Galactic rotation. 
We emphasize that this is only an illustration and is not obtained from a physical model of a propagating gravitational wave. 

Figure~\ref{fig:slab-cepheids} thus shows that nearly all galaxies currently used to calibrate $H_0$ could be located in the contracting part of a large-amplitude, long-wavelength gravitational wave at the current time. 

\begin{figure*}
    \centering
    \includegraphics[width=0.8\textwidth]{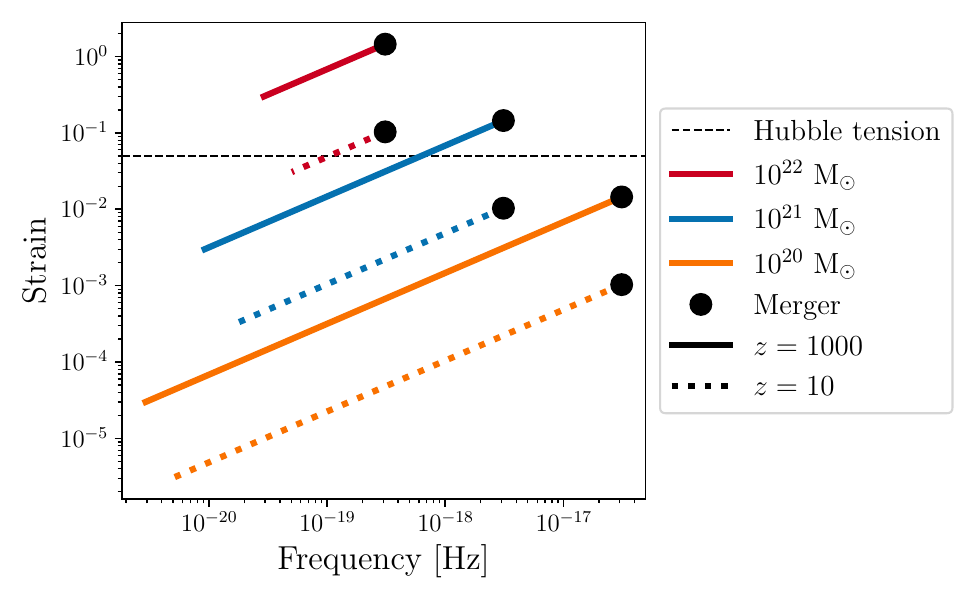}
    \caption{Strain amplitudes of inspiraling stupendously large black holes required to cause the Hubble tension, shown at redshift $z=1000$ (solid lines) and $z=10$ (dotted lines). Note that due to redshift bias, at cosmological distances more distant sources are brighter than closer ones. \label{fig:slab-strain}}
\end{figure*}

\section{Theory}
To estimate the gravitational-wave strain $h$ required to cause a 5\% difference in the Hubble constant and thus explain the Hubble tension, we use a simplified model that links the strain of GWs to the apparent stretching or squeezing of space, and hence to the apparent change in distances used to calculate $H_0$. 

The Hubble constant is a measure of the Universe's expansion rate, defined as the ratio of the recessional velocity $v$ of a galaxy to its distance $d$ from us: 
\begin{equation}
    H_0 = \frac{v}{d}
\end{equation}
A GW passing between us and a distant galaxy can alter the apparent distance to that galaxy. This apparent change in distance, $\Delta d$, can be related to the strain $h$ of the wave and the original distance $d$ as: 
\begin{equation}
    \Delta d \approx h \cdot d
\end{equation}
For $H_0$ to differ by 5\%, this could be interpreted as an apparent change in distances used to calculate $H_0$ by 5\%. Therefore, the required strain can be approximated as simply ${h = \Delta d/d = 0.05}$. 

Assuming a quasi-circular orbit of a SLAB binary, the strain amplitude of inspiraling SLABs at a redshift $z$ and observed frequency $f$ is given by \citep{Maggiore2008GravitationalWaves}: 
\begin{equation}
    h = \frac{4\pi^{2/3} [GM_c(1+z)]^{5/3}}{c^4 D_L(z)} f^{2/3}
    \label{eq:strain}
\end{equation}
%\citep[e.g.,][]{2004ApJ...615...19E}: 
%\begin{eqnarray}
%  h(z,f,M_{\rm{chirp}}) = &3.5 \times 10^{-17} \left(\dfrac{M_{\rm chirp}
%			}{10^8 \ \rm{M}_{\odot}}\right)^{5/3} \nonumber\\ &\times \left[\dfrac{D(z)}{1 \ {\rm Gpc}} \right]^{-1} \left[\dfrac{f (1+z)}{10^{-7} \ {\rm Hz}} \right]^{2/3}
%   \label{eq:strain}
%\end{eqnarray}
where ${M_c}=[M_1 M_2 (M_1 + M_2)^{-1/3}]^{3/5}$ is the chirp mass and $D_L$ the luminosity distance of the SLABs, for which we adopt the \citealt{2020A&A...641A...6P} cosmological parameters. 
Notice in this equation that very high-redshift binaries can have a significantly larger strain than nearby ones \citep{2016PhRvL.116j1102R}. 

The masses required to produce the requisite strain are shown in Figure~\ref{fig:slab-strain}. 
We assume the SLABs to be located at $z=1000$ and compute the inspiral for binary separations ranging from the size of the particle horizon down to the Innermost Stable Circular Orbit $f_{\rm ISCO}=(1/12\pi\sqrt{6})\times c^3/G(M_1+M_2)$, after which the binaries are assumed to merge. 
We find for example that two SLABs with a chirp mass of $10^{21}$~M$_\odot$ at $10^{-18}$~Hz would be sufficient to cause the Hubble tension. 
Such black holes would be located about 10~$R_s$ apart and would still require much longer than the Hubble time to merge \citep{1964PhRv..136.1224P}. %residence time $f/\dot{f}=(5/96)(GM_{\rm chirp}/c^3)^{-5/3}(\pi f)^{-8/3}$, which in this particular example gives 5.47e11 yr and ensures that your binary remains at the same long period.}

Supermassive SLABs may not exist, but they would be big if true: when $10^{22}~\rm{M}_\odot$, their Schwarschild radius is $R_s \simeq 1~\rm{Gpc}$, and so they would likely be the largest objects in existence. 
Given the long inspiral time it is unlikely any such supermassive SLABs have ever merged. 
We dub this the Final Gigaparsec Problem. 

\begin{figure}
\centering
\begin{tikzpicture}
    % Orbits
    \draw[dashed] circle (3cm);
    \draw[dashed] (3,0) circle (0.75cm);

    % Universes 
    \node[circle, fill=customRed!50, minimum size=1.5cm, label=above:The Universe, text=white, label=center:{$10^{24}~\mathrm{M}_\odot$}] (primary) at (0,0) {};
    
    \node[circle, fill=white, minimum size=0.7cm] at (3,0.75) {};
    \node[circle, fill=white, minimum size=0.7cm] at (3,-0.75) {};
    
    \node[circle, fill=customBlue!50, minimum size=0.7cm, label={[align=center, yshift=1mm, xshift=5mm]:Bubble\\ Universes}, label=center:{$10^{22}$}] (secondary1) at (3,0.75) {};
    \node[circle, fill=customBlue!50, minimum size=0.7cm, label=center:{$10^{22}$}] (secondary2) at (3,-0.75) {};
    
    % Gravitational radiation
    \draw[decorate, decoration={snake, segment length=2mm, amplitude=.5mm},->] (3,0) -- +(45:1.5cm);
    \draw[decorate, decoration={snake, segment length=2mm, amplitude=.5mm},->] (3,0) -- +(135:1.5cm);
    \draw[decorate, decoration={snake, segment length=2mm, amplitude=.5mm},->] (3,0) -- +(-45:1.5cm);
    \draw[decorate, decoration={snake, segment length=2mm, amplitude=.5mm},->] (3,0) -- +(-135:1.5cm);
    
    % Adding curved label "Cosmological Horizon"
    \path[
        decorate,
        decoration={
            text along path,
            text={|\sffamily|Cosmological Horizon},
            text align={align=center},
            reverse path
        }
    ] (3cm,0.1) arc (0:-180:3cm);
\end{tikzpicture}
\caption{Schematic of our Universe located in a hierarchical triple system with a pair of inspiraling bubble universes. As gravitational radiation travels at the speed of light, the orbital separation must be at most the cosmological horizon. \label{fig:bubble}} %in order for the gravitational radiation to reach us. \label{fig:bubble}}
\end{figure}
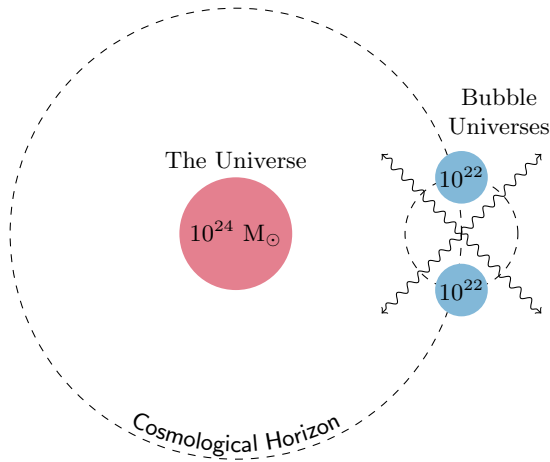

Figure~\ref{fig:slab-strain} further shows the inspiral of two $10^{22}~\rm{M}_\odot$ SLABs. 
We note that this approaches the mass of our Universe. 
Using our adopted cosmological parameters, we estimate the mass of the Universe as product of the volume of the observable Universe and the critical density to find it as approximately $10^{24}~\rm{M}_\odot$. 
Hence such a strain could originate from a pair of inspiraling bubble universes \citep{1984PhLB..136..157G, 2007PhRvD..76f3509A}. We searched the literature for a formula to compute the strain amplitude from coalescing universes, but were unable to find a reference; we thus assume it to also follow Equation~(\ref{eq:strain}). Figure~\ref{fig:bubble} illustrates this idea.

\section{Conclusions}
Inspiraling stupendously large black holes could generate sufficient strain to cause the observed Hubble tension. 
However, this scenario is highly fine-tuned. 
Their gravitational waves would need to propagate along or near the plane of the Milky Way, and their masses would vastly exceed any known black hole mass. 
The formation channel of such supermassive SLABs is unclear, but would likely need to proceed via mergers, as gas accretion is impossible above about $10^{10}~\rm{M}_\odot$. 
However, the merger timescale for such objects greatly exceeds the age of the Universe, and so it is unlikely that such objects could be created in the first place; even the orbital period itself is approximately the Hubble time. 
Furthermore, these objects would need to be located within a few Schwarzschild radii to each other, and it is difficult to even speculate how a pair of such objects could be born so close to one another at the beginning of time. 

Some of these issues can be circumvented by considering the possibility that the sources are formed externally from our Universe and have since entered the cosomological horizon. 
If the Universe were in a triple system with two other coalescing universes located at the horizon, their gravitational radiation could then cause the requisite warping of space. 

This idea could be falsified by observing Cepheid-host galaxies of low galactic latitude: the Hubble constant derived from such observations would show an equal and opposite Hubble tension.

%\acknowledgements{The authors are grateful to the hard-working scientists who had to suffer through lengthy discussions of this extraordinarily implausible thought experiment. Their names have been omitted to protect the innocent.}

%\clearpage
\bibliographystyle{aasjournal.bst}%
\bibliography{main}%

\end{document}